\documentclass[12pt]{article}
\usepackage{graphicx}
\catcode`\@=11
\def\numberbysection{\@addtoreset{equation}{section}
\def\theequation{\thesection.\arabic{equation}}}

\numberbysection

\newcommand{\beq}{\begin{equation}}
\newcommand{\beqa}{\begin{eqnarray}}
\newcommand{\eeq}{\end{equation}}
\newcommand{\eeqa}{\end{eqnarray}}
\newcommand{\abs}[1]{\vert#1\vert}
\newcommand{\bigstat}[1]
{\left\langle\!\!\!\left\langle#1\right\rangle\!\!\!\right\rangle}
\renewcommand{\d}{{\rm d}}
\newcommand{\dstar}{{\star\star}}
\newcommand{\e}{{\rm e}}
\newcommand{\eps}{\varepsilon}
\newcommand{\frad}[2]{\displaystyle{\displaystyle#1\over\displaystyle#2}}
\newcommand\gammaratio{\frac{\Gamma(2/3)}{\Gamma(1/3)}}
\renewcommand{\i}{{\rm i}}
\newcommand{\mean}[1]{\langle#1\rangle}
\newcommand{\stat}[1]{\left\langle\!\left\langle#1\right\rangle\!\right\rangle}
\newcommand{\Ai}{{\rm Ai}}
\newcommand{\Bi}{{\rm Bi}}
\newcommand{\E}{{\cal E}}
\newcommand{\I}{^{\rm I}}
\newcommand{\R}{^{\rm R}}

\begin{document}
\centerline{\Large\bf Non-monotonic disorder-induced enhanced tunneling}
\vspace{1.6cm}
\centerline{\large J.M.~Luck\footnote{luck@spht.saclay.cea.fr}}
\vspace{.4cm}
\centerline{Service de Physique Th\'eorique\footnote{URA 2306 of CNRS},
CEA Saclay, 91191 Gif-sur-Yvette cedex, France}
\vspace{1cm}
\begin{abstract}
The quantum-mechanical transmission through a disordered
tunnel barrier is investigated analytically in the following regime:
(correlation range of the random potential) $\ll$ (penetration length)
$\ll$ (barrier length).
The mean and/or the width of the potential
can either be constant, or vary slowly across the barrier.
The typical transmission is found
to be a non-monotonic function of the disorder strength,
increasing at weak disorder, reaching a maximum in the crossover
from weak to strong disorder, and decreasing at strong disorder.
This work provides a quantitative analysis of the
phenomenon of disorder-induced enhanced tunneling,
put forward by Freilikher et al.
[Phys.~Rev.~E {\bf 51}, 6301 (1995); B {\bf 53}, 7413 (1996)].
\end{abstract}

\vfill
\noindent To be submitted for publication to the Journal of Physics A

\noindent P.A.C.S.: 03.65.Xp, 73.23.-b, 03.65.Nk, 73.20.Fz, 03.65.Sq.

\newpage
\section{Introduction}

The Anderson localization is one of the most spectacular
disorder-induced phenomena~\cite{rev}.
The one-dimensional situation is especially well-understood~\cite{pen}.
Consider for definiteness the Schr\"odinger equation
for an electron moving on a line, in a disordered potential~$V(x)$.
Even in the presence of an infinitesimal amount of disorder,
all eigenstates are exponentially localized,
with a localization length $\xi=1/\gamma$,
where~$\gamma$ is the Lyapunov exponent.
As a consequence, the typical conductance of a disordered sample
falls off exponentially with its length $L$.
More precisely, the zero-temperature conductance $g$ of a one-channel sample
is related to the transmission $T$ across the sample
by the two-probe Landauer formula~\cite{lan}:
\beq
g=\frac{2e^2}{h}\,T.
\label{lanfor}
\eeq
The theory of one-dimensional localization predicts that
the transmission $T$ is a widely fluctuating quantity
in the insulating regime $(\gamma L\gg1)$,
so that the meaningful quantity to consider
is its typical (i.e., most probable) value $\exp(\mean{\ln T})$.
The mean $\mean{\ln T}$ grows linearly with the sample length,
\beq
\mean{\ln T}\approx -2\gamma L,
\label{tintro}
\eeq
and the ratio $(\ln T)/L$ is self-averaging,
in the strong sense that all the cumulants of $(\ln T)$
grow linearly with $L$~\cite{pen}.
In other terms, the statistics of $T$ (resp.~of $\ln T$)
is similar to that of the partition function
(resp.~of the total free energy)
of a disordered thermodynamical system.
This deep analogy appears clearly in the framework of
the transfer-matrix formalism,
especially for discrete (tight-binding) models~\cite{cpv,alea}.

So far, it was implicitly assumed that the energy $E$
of the incoming electron is above the mean of the disordered potential
(usually taken to be zero).
Much less is known on the converse situation
of {\it tunneling through a disordered barrier},
where the mean potential~$\mean{V(x)}$ inside the sample is non-zero,
and higher than the energy~$E$.
More generally, a tunneling situation is met
if the disordered sample is periodic on average,
and if the energy is in a gap
of the underlying average structure~\cite{f1}.

It has been put forward, seemingly in~\cite{f1} for the first time,
that a weak disorder enhances the transmission in such a tunneling situation.
In their subsequent work~\cite{f2},
the authors show that a weak disorder increases
both the mean conductance (proportional to~$\mean{T}$)
and the mean resistance (proportional to~$\mean{1/T}$) of a tunnel barrier.
This disorder-induced enhanced tunneling effect is paradoxical,
because random impurity potentials usually lead to additional scattering,
which hinders transport.
In any case, a strong enough disordered potential is expected to have
the usual effect of reducing the transmission.
Putting together these observations, it can therefore be anticipated
that the transmission reaches a maximum in an intermediate crossover regime,
corresponding to a moderate amount of disorder.
This non-monotonic behavior of the transmission as a function
of the disorder strength seems to have been overlooked so far
(see, however,~\cite{hein} for attempts earlier to~\cite{f1,f2},
and~\cite{dey} for a recent discussion of the effects
of a weak disorder on gap states).

Our aim is to provide a quantitative analysis of the non-monotonic behavior
of the disorder-induced enhanced tunneling transmission.
We restrict this study to the regime of most physical interest:
\beq
a\hbox{ (correlation range of potential)}
\ll 1/K\hbox{ (penetration length)}
\ll L\hbox{ (barrier length)}.
\label{regime}
\eeq
The first inequality implies that the fluctuations of the disordered potential
are short-ranged, so that the latter can be modeled as a Gaussian white noise.
The second inequality implies that the transmission of the barrier
is exponentially small, even in the absence of disorder.
This suggests that $\mean{\ln T}$ will be the right quantity to consider.

For completeness,
we first give in Section~2 an overview of tunneling through a clean barrier.
Section~3 then deals with tunneling through a square disordered barrier
(the mean and the width of the potential are constant),
while Section~4 is devoted to the general case
(the mean and/or the width of the potential vary smoothly across the barrier).
A summary and an outlook are presented in Section~5.

\section{Tunneling through a clean barrier}

We begin with a reminder on the well-known problem
in Quantum Mechanics~\cite{ll} of tunneling through a clean barrier.
In reduced units ($\hbar=2m=1$),
the one-dimensional Schr\"odinger equation reads
\beq
-\psi''(x)+V(x)\psi(x)=E\psi(x).
\label{sch}
\eeq

\subsubsection*{Square clean barrier}

Consider first the simple case of a square barrier of length $L$.
The potential is constant in the barrier:
\beq
V(x)=V_0\qquad(0\le x\le L),
\eeq
and vanishes elsewhere.

In the situation of interest,
the energy $E$ of the incoming particle is in the range $0<E<V_0$.
The wavevector $p$ of the particle
and its inverse penetration length $K$ in the barrier read
\beq
p=\sqrt{E},\qquad K=\sqrt{V_0-E}.
\label{defpk}
\eeq
The reflection and transmission amplitudes $r$ and $t$
are determined by looking for a solution to~(\ref{sch}) of the form
\beq
\psi(x)=\left\{\matrix{
\e^{\i px}+r\,\e^{-\i px}\hfill&&(x\le0),\hfill\cr
a\,\e^{Kx}+b\,\e^{-Kx}\hfill&&(0\le x\le L),\hfill\cr
t\,\e^{\i p(x-L)}\hfill&&(x\ge L).\hfill
}\right.
\label{wave}
\eeq
Expressing the continuity of $\psi(x)$ and of its derivative
at $x=0$ and $x=L$ provides four linear equations, whose solution yields
\beqa
&&r=\frad{(p^2+K^2)\sinh KL}{(p^2-K^2)\sinh KL+2\i pK\cosh KL},\label{r}\\
&&t=\frad{2\i pK}{(p^2-K^2)\sinh KL+2\i pK\cosh KL}.\label{t}
\eeqa

Throughout the following, we will be mostly interested in
the transmission intensity coefficient (or transmission for short),
\beq
T=\abs{t}^2,
\eeq
which enters the Landauer formula~(\ref{lanfor}).
In the present case,~(\ref{t}) yields
\beq
T=\frac{4p^2K^2}{4p^2K^2+(p^2+K^2)^2\sinh^2KL}.
\eeq
In the regime~(\ref{regime}),
where the barrier length is much larger than the penetration length,
the transmission falls off exponentially, as
\beq
T\approx\frac{16p^2K^2}{(p^2+K^2)^2}\,\exp(-2KL).
\label{tcarreasy}
\eeq
All subsequent results for the transmission will be given
{\it with exponential accuracy}, in analogy with~(\ref{tintro}).
Neglecting the prefactor, we thus rewrite~(\ref{tcarreasy})~as
\beq
\ln T\approx-2KL.
\label{tcarre}
\eeq

\subsubsection*{Arbitrary clean barrier}

Let us now consider tunneling through an arbitrary clean barrier.
The potential $V(x)$ is larger than the energy $E$ for $0\le x\le L$,
so that the inverse penetration length reads
\beq
K(x)=\sqrt{V(x)-E}.
\eeq
We assume that the potential has a smooth profile across the barrier,
i.e., the length scale over which $V(x)$ or $K(x)$ varies
is of the order of the barrier length $L$ itself.

The transmission can be determined along the lines of the previous case,
by seeking a solution to~(\ref{sch}) of the form~(\ref{wave}),
where $\exp(\pm Kx)$ are replaced by the two elementary
solutions $u(x)$ and $v(x)$,
with initial values $u(0)=v'(0)=1$, $u'(0)=v(0)=0$,
whose Wronskian reads $u(x)v'(x)-u'(x)v(x)=1$.
We thus obtain
\beq
t=\frac{2\i p}{p^2v(L)+\i p(u(L)+v'(L))-u'(L)}.
\label{tuv}
\eeq

The hypothesis of a smoothly varying potential implies that it is legitimate
to use the well-known W.K.B. approximation~\cite{ll,wkb}.
Indeed, the condition for this scheme to be valid,
\beq
\frac{1}{K(x)^2}\,\frac{\d K(x)}{\d x}\sim \frac{1}{K(x)L}\ll1,
\eeq
is automatically satisfied in the regime~(\ref{regime}) of long barriers.
Within this framework, a basis of solutions to~(\ref{sch}) in the barrier reads
\beq
\psi_\pm(x)\sim\exp\left(\pm\int_0^x K(y)\,\d y\right)\qquad(0<x<L).
\label{psiwkb}
\eeq
At least one of the elementary solutions $u(x)$ and $v(x)$
(and generically both of them)
is proportional to the growing solution $\psi_+(x)$.
Hence~(\ref{tuv}) leads to the estimate
\beq
T\sim\frac{1}{\abs{\psi_+(L)}^2},
\label{tsim}
\eeq
which will be sufficient hereafter, in order to work with exponential accuracy.

The transmission therefore again falls off exponentially:
\beq
\ln T\approx-2\int_0^L K(x)\,\d x,
\label{twkb}
\eeq
where the integral is the action of the classical imaginary-time trajectory,
or {\it instanton}, crossing the barrier at energy $E$~\cite{ll,wkb,jzj}.
The estimate~(\ref{twkb}) generalizes~(\ref{tcarre}) to a potential barrier
with an arbitrary (smooth) profile.

\section{Tunneling through a square disordered barrier}

We now turn to the case of a square disordered barrier of length $L$.
The barrier potential,
\beq
V(x)=V_0+W(x)\qquad(0\le x\le L),
\eeq
is the sum of a constant $V_0$ and of a disordered component $W(x)$
with zero mean.

As stated in the Introduction, in the regime~(\ref{regime})
it is legitimate to model $W(x)$ as a Gaussian white noise, such that
\beq
\mean{W(x)W(y)}=2D\,\delta(x-y).
\label{ww}
\eeq
The estimate~(\ref{tsim}) still holds in the presence of disorder.
We are therefore led to consider the situation where the random potential
extends over the half-line $x>0$, and to investigate the growth rate
of the generic solution to~(\ref{sch}).

The most salient effect of the disordered potential
is that the wavefunction in the barrier now changes sign many times
in the regime under consideration,
so that neither the concept of a single instanton trajectory,
nor the W.K.B. estimate~(\ref{psiwkb}), make sense anymore.
The most efficient approach to this problem is the invariant-measure method,
initiated long ago by Dyson~\cite{dy} and Schmidt~\cite{sch}.
For technical reasons,
the energy $E-\i0$ and the inverse penetration length $K+\i0$
are respectively endowed
with infinitesimal negative and positive imaginary parts.
The method consists in introducing the Riccati variable
\beq
z(x)=\frac{\psi'(x)}{\psi(x)},
\label{zdef}
\eeq
so that
\beq
\psi(x)=\psi(0)\exp\int_0^x z(y)\,\d y.
\label{psiexp}
\eeq

The Schr\"odinger equation~(\ref{sch}) is equivalent to the Riccati equation
\beq
z'=K^2-z^2+W(x).
\label{ric}
\eeq
The key property of this equation is the following:
the complex random variable $z(x)$ has a well-defined
limit probability distribution, irrespective of the position $x$,
provided it is deep enough in the sample $(Kx\gg1)$,
and of the initial condition $z(0)$.
Let us denote averages with respect to this invariant measure as $\stat{...}$.
Equation~(\ref{psiexp}) implies that the growing solution to~(\ref{sch})
typically grows exponentially, as
\beq
\ln\psi_+(x)\approx\Omega x,
\label{omedef}
\eeq
where
\beq
\Omega=\stat{z}
\eeq
is the complex characteristic exponent.
When the energy variable is just below the real axis,
$\Omega$ splits according to~\cite{tn,alea}
\beq
\Omega(E-\i0)=\gamma(E)+\i\pi H(E).
\label{split}
\eeq

The real part of~(\ref{split}) is the Lyapunov exponent $\gamma$.
Inserting the behavior~(\ref{omedef}) into the estimate~(\ref{tsim}),
we therefore obtain at once the prediction
\beq
\mean{\ln T}\approx -2\gamma L,
\label{tdis}
\eeq
recalled in the Introduction~[see~(\ref{tintro})].
The imaginary part of~(\ref{split}) is proportional
to the integrated density of states of the problem per unit length,
\beq
H(E)=\int_{-\infty}^E\rho(E')\,\d E',
\eeq
so that $1/H(E)$ is the mean distance between any two consecutive zeros
of $\psi_+(x)$.

The present case of a white-noise potential has been investigated
by several authors~\cite{fl,h,sulem}.
Their findings can be recast into the following
formula for the characteristic exponent:
\beq
\Omega=D^{1/3}\,F(X),
\label{char}
\eeq
with
\beq
X=\frac{K^2}{D^{2/3}}=\frac{V_0-E}{D^{2/3}}
\label{xdef}
\eeq
and
\beq
F(X)=\e^{-2\i\pi/3}\frac{\Ai'(\e^{-2\i\pi/3}X)}{\Ai(\e^{-2\i\pi/3}X)}
=\frac{\Ai'(X)+\i\,\Bi'(X)}{\Ai(X)+\i\,\Bi(X)},
\label{fdef}
\eeq
where $\Ai(z)$ and $\Bi(z)$ are Airy functions~\cite{as}.
As a consequence of~(\ref{split}), we have
\beqa
&&{\hskip 1.4mm}\gamma=D^{1/3}\,F\R(X),
{\hskip 8.6mm}F\R(X)=
\frac{\Ai(X)\Ai'(X)+\Bi(X)\Bi'(X)}{\Ai(X)^2+\Bi(X)^2},\label{charr}\\
&&H=\frac{D^{1/3}}{\pi}\,F\I(X),
{\hskip 10mm}F\I(X)=\frac{1}{\pi\left(\Ai(X)^2+\Bi(X)^2\right)}.
\label{chari}
\eeqa
References~\cite{tn,fl,h,sulem} deal separately
with the real and imaginary parts of $\Omega$,
and therefore rather derive~(\ref{charr}) and/or~(\ref{chari}).
For completeness, we give in the Appendix
a self-contained derivation of~(\ref{char})--(\ref{fdef}).

Inserting~(\ref{charr}) into~(\ref{tdis}) leads us to the prediction
\beq
\mean{\ln T}\approx -2D^{1/3}\,F\R(X)\,L,
\eeq
where the scaling variable $X$ is real and positive
in a tunneling situation.

In order to emphasize the dependence of the transmission on the disorder
strength, we recast the above prediction as
\beq
\mean{\ln T}\approx(\ln T)_0\,G(Y),
\label{main}
\eeq
where $(\ln T)_0=-2KL$ is the result~(\ref{tcarre})
in the absence of disorder, while
\beq
Y=\frac{D}{K^3}=\frac{D}{(V_0-E)^{3/2}}=X^{-3/2}
\label{ydef}
\eeq
is the reduced disorder strength, and the scaling function~$G$ reads
\beq
G(Y)=Y^{1/3}\,F\R(Y^{-2/3}).
\label{gdef}
\eeq

\subsubsection*{Weak-disorder regime}

The weak-disorder regime corresponds to $D\ll K^3$, i.e., $X\gg1$ or $Y\ll1$.
The differential equation
\beq
F^2+F'=X
\eeq
obeyed by the function $F(X)$ easily yields the asymptotic expansion
\beq
F(X)=X^{1/2}-\frac{1}{4X}-\frac{5}{32X^{5/2}}+\cdots,
\eeq
This result is formally real, so that $F\R$ has the same asymptotic expansion
(while $F\I$ is exponentially small as $X\to+\infty$).
We thus obtain
\beq
G(Y)=1-\frac{Y}{4}-\frac{5Y^2}{32}+\cdots,
\label{gweak}
\eeq
i.e., more explicitly
\beq
\mean{\ln T}\approx2L\left(-K+\frac{D}{4K^2}+\frac{5D^2}{32K^5}\cdots\right).
\label{tweak}
\eeq
The first correction of order $D$
is in agreement with the perturbative result of~\cite{f2}:
the leading effect of a weak disorder is found to enhance transmission.

\subsubsection*{Strong-disorder regime}

Consider now the opposite regime of strong disorder
($D\gg K^3$, i.e., $Y\gg1$ or $X\ll1$).
In this regime, the scaling function
\beq
G(Y)\approx F\R(0)\,Y^{1/3}
\label{gstrong}
\eeq
grows as the power $1/3$ of the strength of disorder,
with the explicit prefactor
\beq
F\R(0)=\frac{3^{1/3}}{2}\,\gammaratio=0.364506.
\eeq
As a consequence, we have
\beq
\mean{\ln T}\approx-2F\R(0)\,D^{1/3}\,L.
\label{tstrong}
\eeq
The transmission decreases with the disorder strength
in the strong-disorder regime, as anticipated in the Introduction.
The result~(\ref{tstrong}) is independent of $K$:
the damping due to the mean barrier $V_0$
becomes negligible with respect to localization effects.

\subsubsection*{Non-monotonic crossover behavior}

The function $G(Y)$ which enters the scaling law~(\ref{main})
decreases first from its value $Y(0)=1$, according to~(\ref{gweak}),
in the weak-disorder regime,
and then increases according to~(\ref{gstrong}), in the strong-disorder regime.
It must therefore reach a minimum,
somewhere in the crossover from weak to strong disorder.

\begin{figure}[ht]
\begin{center}
\includegraphics[angle=90,width=.495\linewidth]{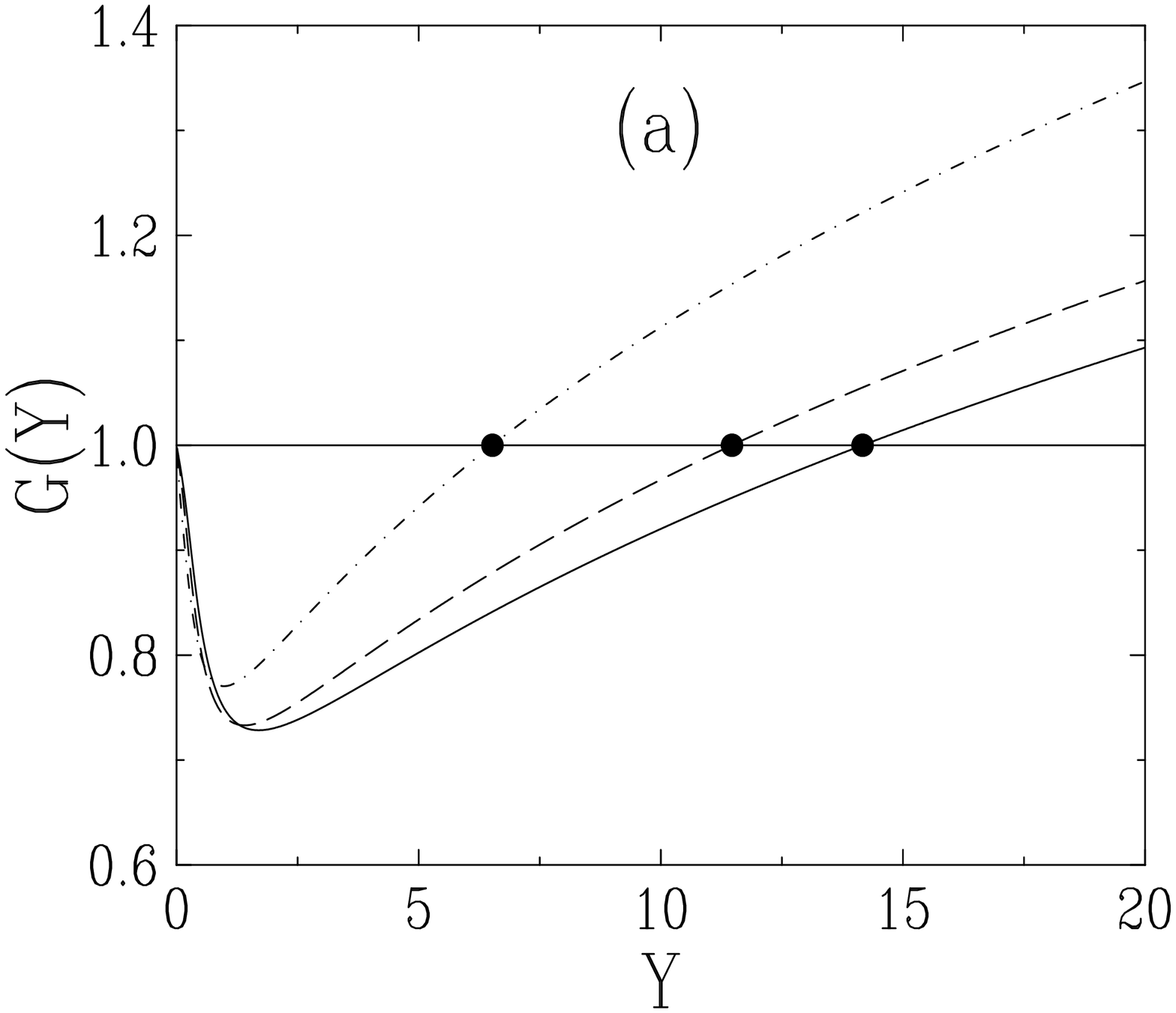}
\includegraphics[angle=90,width=.49\linewidth]{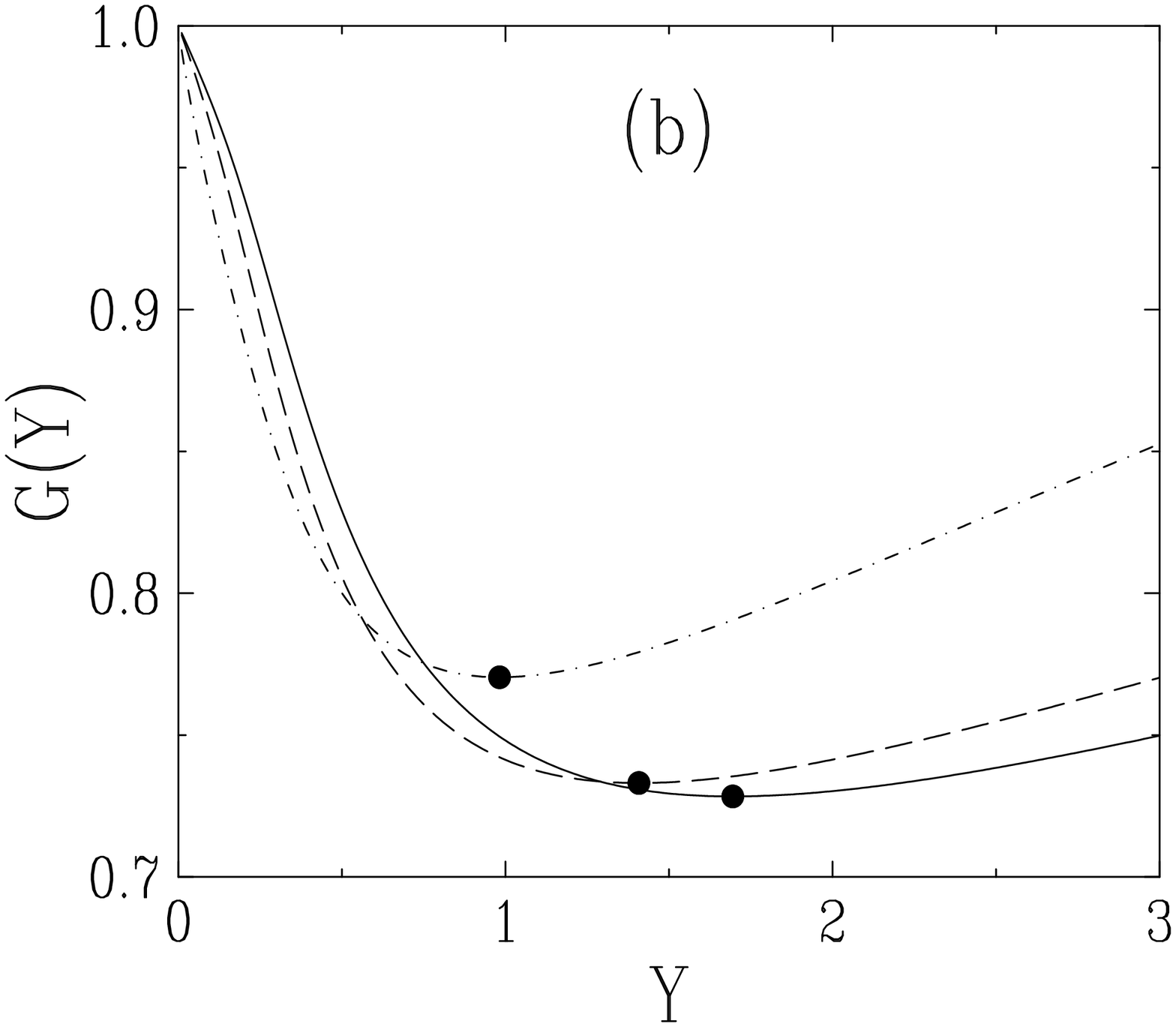}
\caption{\small
Plots of the scaling functions describing the effect of disorder
on tunneling transmission, against the reduced disorder strength $Y$.
Full lines: $G$ entering~(\ref{main}) (square barrier).
Dashed lines: $G_1$ entering~(\ref{exa1})
(parabolic barrier with parabolic disorder).
Dash-dotted lines: $G_2$ entering~(\ref{exa2})
(parabolic barrier with uniform disorder).
(a): Circles show the values of $Y$ where the scaling functions cross unity:
$Y^\dstar=14.168$, $Y_1^\dstar=11.469$, $Y_2^\dstar=6.522$.
(b) (enlargement): Circles show the values of $Y$ where the
scaling functions have their minima:
$G=0.7284$ for $Y^\star=1.695$,
$G_1=0.7331$ for $Y_1^\star=1.408$,
$G_2=0.7704$ for $Y_2^\star=0.982$.}
\label{fig1}
\end{center}
\end{figure}

Figure~\ref{fig1} shows plots (full lines) of the scaling function $G(Y)$,
given by~(\ref{charr}), (\ref{gdef}).
This function passes through a minimum, $G(Y^\star)=0.7284$ for $Y^\star=1.695$,
before it crosses the value $G(Y^\dstar)=1$ for $Y^\dstar=14.168$.
The non-monotonic behavior of the transmission as a function
of the disorder strength in the crossover regime,
anticipated in the Introduction,
is therefore confirmed at a quantitative level.
The tunneling transmission is enhanced for a weak enough
reduced disorder strength ($Y<Y^\dstar$),
the enhancement being maximal for~$Y=Y^\star$.

\section{Tunneling through an arbitrary disordered barrier}

We finally turn to the general situation
where both the deterministic part $V_0(x)$
and the strength of the disordered part $W(x)$ of the potential
may have a smooth dependence on the position across the barrier.
We set
\beq
V(x)=V_0(x)+W(x)\qquad(0<x<L),
\eeq
with
\beq
K(x)=\sqrt{V_0(x)-E},
\eeq
and
\beq
\mean{W(x)W(y)}=2D(x)\,\delta(x-y).
\eeq

We assume that $K(x)$ and $D(x)$ vary smoothly across the barrier,
and we again focus our attention onto the regime~(\ref{regime}).
In this situation, the Riccati variable $z(x)$
is approximately distributed according to the local invariant measure,
characterized by the parameters $K(x)$ and $D(x)$.
The reason why this adiabatic approach is legitimate
is the same as for the W.K.B. scheme in the absence of disorder:
the length over which parameters vary, i.e., the barrier length~$L$ itself,
is much larger than the characteristic length
of the relaxation of the distribution of the Riccati variable,
i.e., the localization length $1/\gamma(x)$.

Equations~(\ref{tsim}), (\ref{psiexp}) yield the following expression
\beq
\mean{\ln T}\approx-2\int_0^L\gamma(x)\,\d x
\label{twd}
\eeq
for the mean logarithm of the transmission.
The prediction~(\ref{twd}) includes all the previous results~(\ref{tcarre}),
(\ref{twkb}), (\ref{tdis}) as special cases.
By means of~(\ref{charr}), it can be recast into the more explicit form
\beq
\mean{\ln T}\approx-2\int_0^L
D(x)^{1/3}\,F\R\!\left(\frac{K(x)^2}{D(x)^{2/3}}\right)\d x.
\label{texpli}
\eeq

In the weak-disorder regime, this prediction behaves as
\beq
\mean{\ln T}\approx2\int_0^L\left(-K(x)+\frac{D(x)}{4K(x)^2}+\cdots\right)\d x,
\eeq
while in the strong-disorder regime we have
\beq
\mean{\ln T}\approx-2F\R(0)\int_0^LD(x)^{1/3}\,\d x.
\eeq
These two expressions respectively generalize~(\ref{tweak})
and~(\ref{tstrong}).

To close up, let us illustrate more explicitly the result~(\ref{texpli})
on two examples.

\subsubsection*{Example~1: Parabolic barrier with parabolic disorder}

In our first example, both the deterministic potential
and the disorder strength have a parabolic shape:
\beq
K(x)^2=V_0(x)-E=K_0^2\,\frac{4x(L-x)}{L^2},\qquad
D(x)=D_0\,\frac{4x(L-x)}{L^2},
\label{para}
\eeq
with maximum values $K_0^2$ and $D_0$ at the center of the barrier ($x=L/2$).

In this situation,~(\ref{texpli}) yields the scaling law
\beq
\mean{\ln T}\approx(\ln T)_0\,G_1(Y),
\label{exa1}
\eeq
similar to~(\ref{main}), with
\beq
(\ln T)_0=-\frac{\pi}{2}\,K_0L,
\label{tvar}
\eeq
\beq
Y=\frac{D_0}{K_0^3},
\eeq
and
\beq
G_1(Y)=\frac{2}{\pi}\,Y^{1/3}\int_0^\pi
F\R\left(Y^{-2/3}(\sin\theta)^{2/3}\right)(\sin\theta)^{5/3}\,\d\theta.
\label{g1def}
\eeq
The latter expression is obtained by setting $x=L(1+\cos\theta)/2$,
so that $4x(L-x)/L^2=\sin^2\theta$, with $0\le\theta\le\pi$.

The scaling function $G_1(Y)$ has been evaluated numerically by means of the
integral~(\ref{g1def}), and plotted in Figure~\ref{fig1} (dashed lines).
Its qualitative dependence on the disorder strength $Y$
is similar to that of $G(Y)$,
with the following behavior at weak and strong disorder:
\beq
\matrix{
G_1(Y)\approx1-\frad{Y}{\pi}\hfill&(Y\ll1),\cr\cr
G_1(Y)\approx\frad{12^{1/3}\Gamma(1/3)}{5\pi}\,Y^{1/3}\approx0.390454\,Y^{1/3}
\qquad&(Y\gg1),}
\eeq
with a minimum $G_1(Y_1^\star)=0.7331$ for $Y_1^\star=1.408$,
and with $G_1(Y_1^\dstar)=1$ for $Y_1^\dstar=11.469$.

\subsubsection*{Example~2: Parabolic barrier with uniform disorder}

In our second example, the deterministic potential still has
the parabolic form~(\ref{para}),
while the disorder strength $D(x)=D$ is constant.

The prediction~(\ref{texpli}) now yields the scaling form
\beq
\mean{\ln T}\approx(\ln T)_0\,G_2(Y),
\label{exa2}
\eeq
again with~(\ref{tvar}), and with
\beq
Y=\frac{D}{K_0^3},
\eeq
and
\beq
G_2(Y)=\frac{2}{\pi}\,Y^{1/3}\int_0^\pi
F\R\left(Y^{-2/3}\sin^2\theta\right)\sin\theta\,\d\theta.
\eeq

The scaling function $G_2(Y)$ is also plotted in Figure~\ref{fig1}
(dash-dotted lines).
Its qualitative dependence on the disorder strength $Y$
is again similar to that of $G(Y)$,
with the following behavior at weak and strong disorder:
\beq
\matrix{
G_2(Y)\approx1-\frad{Y}{3\pi}\ln\frad{1}{Y}\hfill&(Y\ll1),\cr\cr
G_2(Y)\approx\frad{2\,3^{1/3}}{\pi}\gammaratio\,Y^{1/3}\approx0.464103\,Y^{1/3}
\qquad&(Y\gg1),}
\eeq
with a minimum $G_2(Y_2^\star)=0.7704$ for $Y_2^\star=0.982$,
and with $G_2(Y_2^\dstar)=1$ for $Y_2^\dstar=6.522$.

\section{Discussion}

In this paper, we have investigated by analytical means
the transmission through a disordered tunnel barrier,
in the regime~(\ref{regime}) of most physical interest.
We have thus provided a quantitative analysis of the
phenomenon of disorder-induced enhanced tunneling,
put forward by Freilikher et al.~\cite{f1,f2}.
The most salient outcome of the present work
is that the enhancement effect is a non-monotonic
function of the strength of disorder, and that it is maximally efficient
at some well-defined intermediate value $Y^\star$
of the reduced disorder strength~$Y$.

The key point of our approach consists in utilizing
the scaling law~(\ref{char})--(\ref{fdef})
for the complex characteristic exponent $\Omega$.
It is worth noticing that this formalism encompasses,
and treats on the same footing,
both the usual situation of the Anderson localization,
where the energy is above the mean of the disordered potential
(corresponding to negative values of the scaling variable $X$),
and the tunneling situation,
where the energy is below the mean of the disordered potential
(corresponding to positive values of~$X$).
As a consequence, the general results on the statistics of the transmission,
recalled in the Introduction, still hold in the tunneling situation.
In particular, $\mean{\ln T}$ is the right quantity to consider.

In the case of a white-noise potential, considered throughout this work,
the form~(\ref{xdef}) of the scaling variable $X$
is merely dictated by dimensional analysis,
while the explicit formula~(\ref{fdef}) for the scaling function $F(X)$
is given a self-contained derivation in the Appendix.
Somewhat equivalent results had been obtained
in several earlier works~\cite{fl,h,sulem}.
The scaling law~(\ref{char})--(\ref{fdef})
also describes the spectra of other one-dimensional disordered systems,
such as the diffusion of classical particles in a random force field~\cite{bg}.
An analogous scaling formula holds
for discrete models near their band edges.
Consider the tight-binding equation
\beq
\psi_{n-1}+\psi_{n+1}+V_n\psi_n=\E\psi_n.
\label{tbm}
\eeq
For a clean chain ($V_n=0$), the dispersion relation reads $\E=2\cos p$.
In the regime of a weak disorder ($\mean{V_n^2}=\Delta\ll1$),
and near the upper band edge ($p\ll1$), one has~\cite{dg}
\beq
\Omega=\Delta^{1/3}\,f(x),\qquad
x=-\frac{p^2}{\Delta^{2/3}}\approx\frac{\E-2}{\Delta^{2/3}}.
\label{dgres}
\eeq
This scaling law turns out to play a central role in various problems,
such as the spreading dynamics of a wave packet~\cite{toro}.
The results~(\ref{char})--(\ref{fdef})
can be viewed as the formal continuum limit of~(\ref{dgres}).
The identification between our continuum problem
and the tight-binding model~(\ref{tbm}) has to take place near the band edge.
Indeed, introducing explicitly the lattice spacing $a$,
which plays the role of the correlation range of the potential,
the first inequality of~(\ref{regime}) implies $\abs{pa}=Ka\ll1$,
as $p=\i K$, whereas $D=\Delta/(2a^3)$.
The scaling variables and functions match between~(\ref{dgres})
and~(\ref{char})--(\ref{fdef})
(up to powers of $2$ due to different conventions):~$x=2^{2/3}X$, $f=2^{1/3}F$.

For a square disordered barrier,
the main result~(\ref{main}) for $\mean{\ln T}$,
i.e., the logarithm of the typical transmission,
can be generalized to all the moments of $T$ in the regime~(\ref{regime}).
It can indeed be shown, along the lines of~\cite{pen},
that these moments scale as
\beq
\mean{T^n}\approx\exp\left(-2n\,G_n(Y)\,KL\right),
\eeq
where $Y$ is the reduced disorder strength of~(\ref{ydef}),
while the scaling function $G_n(Y)$ depends on the order $n$
(not necessarily an integer).
Equation~(\ref{main}) is recovered as $G_0(Y)=G(Y)$.
Skipping every detail, let us mention the weak-disorder expansion
\beq
G_n(Y)=1-\frac{2n+1}{4}\,Y+\cdots\qquad(Y\ll1),
\eeq
hence
\beq
\mean{T^n}\approx
\exp\left[2L\left(-nK+\frac{n(2n+1)D}{4K^2}+\cdots\right)\right].
\eeq
In agreement with~\cite{f2}, both $\mean{T}$ and $\mean{1/T}$,
and indeed all the moments whose order is not in the range $-1/2<n<0$,
are increased by a weak amount of disorder.
In the opposite strong-disorder regime, we have
\beq
G_n(Y)\approx a_n\,Y^{1/3}\qquad(Y\gg1),
\eeq
with some $n$-dependent amplitude $a_n$, hence
\beq
\mean{T^n}\approx\exp\left(-2n\,a_n\,D^{1/3}L\right).
\eeq
The evaluation of the full scaling function $G_n(Y)$ is a difficult task
in general, except for~$n$ a negative integer,
corresponding by means of~(\ref{tsim}) to positive integer
powers of $\abs{\psi_+(L)}^2$.
In this case, $G_n(Y)$ can be derived
by means of the algebraic approach of~\cite{pen}:
it is (the real part of) an algebraic function with degree $2\abs{n}+1$.

For a disordered barrier with an arbitrary profile,
the general prediction~(\ref{texpli}) has been illustrated on two
realistic examples of parabolic barriers.
These examples demonstrate that the qualitative features of
disorder-induced enhanced tunneling, and chiefly its non-monotonic behavior
as a function of the disorder strength,
are rather insensitive to the shape of the mean and/or the width of the
random potential.
Even quantitative characteristics, such as $Y^\star$ or $Y^\dstar$,
do not depend too much on the profile of the barrier.

Besides the present situation of thick tunnel barriers,
the non-monotonic enhancement of transmission may be a more general phenomenon.
Somewhat similar features have indeed been observed recently~\cite{g}
in a problem inspired by nuclear fission.

It is also worth noticing that,
in the somewhat dual case of the total reflection
of an electron by a semi-infinite disordered sample,
universal features of the distribution of the Wigner time delay
has received much attention recently~\cite{tc}.
The effects of disorder-induced enhanced tunneling
on the Wigner time delay are also potentially of interest
in transmission~\cite{bolt}, even though this concept has been criticized
as being somewhat ambiguous~\cite{dwell}.

Finally, the disorder-induced enhancement of transmission
is a quantum-mechanical phenomenon, and more generally a wave phenomenon.
It is indeed due to the existence of non-trivial localized states
in a weakly-disordered barrier,
which themselves originate in the interferences between the multiply
scattered waves.
Therefore, no disorder-induced enhancement is to be expected
e.g. in the Kramers problem~\cite{mel} of the thermally activated hopping
of a classical particle over a potential barrier.

\subsubsection*{Acknowledgements}

It is a pleasure to thank Bertrand Giraud for discussions related
to~\cite{g} which motivated this work,
and Boris Shapiro for valuable correspondence.

\appendix

\section{Derivation of expressions~(\ref{char})--(\ref{fdef})}

This Appendix presents a self-contained derivation
of expressions~(\ref{char})--(\ref{fdef})
for the complex characteristic exponent $\Omega$, introduced in~(\ref{omedef}).

To do so, we determine the invariant measure
of the complex Riccati variable $z$, introduced in~(\ref{zdef}).
As this distribution has a complex support,
instead of writing a Fokker-Planck equation for its density,
it is advantageous to consider the linear transforms
\beq
\Phi(y)=\stat{\ln(y-z)},\qquad\phi(y)=\Phi'(y)=\bigstat{\frac{1}{y-z}}.
\eeq

For definiteness, we assume that $K$ has positive real and imaginary parts.
Equation~(\ref{ric}) shows that $z(x)$ keeps a positive imaginary part,
so that $\Phi(y)$ is analytic in the lower half plane.

In the spirit of the derivation of the Fokker-Planck equation~\cite{vank},
we consider a small increment $\eps$ of $x$, and set $z(x+\eps)=z(x)+\eta$.
Both
\beq
\mean{\eta}\approx(K^2-z^2)\eps,\qquad\mean{\eta^2}\approx2D\eps
\label{cums}
\eeq
are proportional to the increment $\eps$,
while higher cumulants are negligible.

Along the lines of the Dyson-Schmidt approach~\cite{dy,sch},
we then look for a stationarity condition by comparing
the expressions of $\Phi(y)$ corresponding
to the points $x$ and $x+\eps$: $\stat{\ln(y-z)}=\stat{\ln(y-z-\eta)}$.
By expanding the right side of this equality in powers of $\eta$,
and using~(\ref{cums}), we obtain the condition
\beq
\bigstat{\frac{K^2-z^2}{y-z}+\frac{D}{(y-z)^2}}=0,
\eeq
i.e.,
\beq
D\phi'(y)+(y^2-K^2)\phi(y)=y+\Omega.
\eeq
The strength of disorder $D$ can be scaled out by setting
$K^2=D^{2/3}X$, $\Omega=D^{1/3}F$,
proving thus the scaling formulas~(\ref{char}), (\ref{xdef}),
and $y=D^{1/3}u$, $\phi=D^{-1/3}\psi$.

We are thus left with a differential equation for $\psi(u)$,
\beq
\psi'(u)+(u^2-X)\psi(u)=u+F,
\label{ode}
\eeq
which can be solved by {\it varying the constant}:
\beq
\psi(u)=\exp(-u^3/3+Xu)\,C(u),\qquad C(u)=\int(v+F)\,\exp(v^3/3-Xv)\,\d v.
\eeq
The existence of a regular solution,
such that $\psi(u)\to0$ as $\abs{u}\to\infty$, determines $F(X)$.
We must have $C(u)\to0$ as $\abs{u}\to\infty$
in all the directions of the lower half plane where $\exp(-u^3/3)$ diverges.
This happens in two Stokes sectors, represented by the directions
$u\to-\infty$ and $u\to\e^{-\i\pi/3}\infty$.
We thus obtain
\beq
F=\frac{I'(X)}{I(X)},
\eeq
with
\beq
I(X)=\int_{-\infty}^{\e^{-\i\pi/3}\infty}\exp(v^3/3-Xv)\,\d v.
\eeq
Finally, $I(X)$ is equal to the Airy function $\Ai(\e^{-2\i\pi/3}X)$~\cite{as},
up to a multiplicative constant.
This observation leads to~(\ref{fdef}).


\newpage
\begin{thebibliography}{99}

\bibitem{rev}
See e.g. I.M. Lifshitz, S.A. Gredeskul, and L.A. Pastur, {\it Introduction to
the Theory of Disordered Systems} (Wiley, New-York, 1988);
B. Kramer and A. MacKinnon, Rep. Prog. Phys. {\bf 56}, 1469 (1993);
Y. Imry, {\it Introduction to Mesoscopic Physics} (Oxford University Press,
Oxford, 1997);
K. Efetov, {\it Supersymmetry in Disorder and Chaos} (Cambridge University
Press, Cambridge, 1997).

\bibitem{pen}
J.B. Pendry, Adv. Phys. {\bf 43}, 461 (1994).

\bibitem{lan}
Y. Imry and R. Landauer, Rev. Mod. Phys. {\bf 71}, S306 (1999),
and the references therein.

\bibitem{cpv}
A. Crisanti, G. Paladin, and A. Vulpiani, {\it Products of Random Matrices in
Statistical Physics} (Springer, Berlin, 1992).

\bibitem{alea}
J.M. Luck, {\it Syst\`emes d\'esordonn\'es unidimensionnels} (in French)
(Collection Al\'ea-Saclay, 1992).

\bibitem{f1}
V.D. Freilikher, B.A. Liansky, I.V. Yurkevich, A.A. Maradudin, and A.R. McGurn,
Phys. Rev. E {\bf 51}, 6301 (1995).

\bibitem{f2}
V.D. Freilikher, M. Pustilnik, and I.V. Yurkevich, Phys. Rev. B {\bf 53}, 7413
(1996).

\bibitem{hein}
J. Heinrichs, Phys. Rev. B {\bf 33}, 5261 (1986); {\bf 36}, 2867 (1987).

\bibitem{dey}
L.I. Deych, A.A. Lisyansky, and B.L. Altshuler, Phys. Rev. Lett. {\bf 84}, 2678
(2000).

\bibitem{ll}
See e.g. L.D. Landau and E.M. Lifshitz, {\it Quantum Mechanics:
Non-Relativistic Theory} (Pergamon, Oxford, 1959).

\bibitem{wkb}
See e.g. N. Fr\"oman and P.O. Fr\"oman, {\it Physical Problems Solved by the
Phase-Integral Method} (Cambridge University Press, Cambridge, 2002).

\bibitem{jzj}
See e.g. J. Zinn-Justin, {\it Quantum Field Theory and Critical Phenomena}
(Clarendon, Oxford, 1989).

\bibitem{dy}
F.J. Dyson, Phys. Rev. {\bf 92}, 1331 (1953).

\bibitem{sch}
H. Schmidt, Phys. Rev. {\bf 105}, 425 (1957).

\bibitem{tn}
Th.M. Nieuwenhuizen, Physica {\bf A 120}, 468 (1983).

\bibitem{fl}
H.L. Frisch and S.P. Lloyd, Phys. Rev. {\bf 120}, 1175 (1960).

\bibitem{h}
B.I. Halperin, Phys. Rev. {\bf 139}, A 104 (1965).

\bibitem{sulem}
P.L. Sulem, Physica {\bf 70}, 190 (1973).

\bibitem{as}
M. Abramowitz and I.A. Stegun, {\it Handbook of Mathematical Functions} (Dover,
New York, 1974).

\bibitem{bg}
J.P. Bouchaud, A. Comtet, A. Georges, and P. Le Doussal, Ann. Phys. {\bf 201},
285 (1990).

\bibitem{dg}
B. Derrida and E. Gardner, J. Phys. (France) {\bf 45}, 1283 (1984).

\bibitem{toro}
S. De Toro Arias and J.M. Luck, J. Phys. A {\bf 31}, 7699 (1998).

\bibitem{g}
B.G. Giraud, K. Amos, S. Karataglidis, and B.A. Robson (work in progress).

\bibitem{tc}
C. Texier and A. Comtet, Phys. Rev. Lett. {\bf 82}, 4220 (1999),
and the references therein.

\bibitem{bolt}
C.J. Bolton-Heaton, C.J. Lambert, V.I. Fal'ko, V. Progodin, and A.J. Epstein,
Phys. Rev. B {\bf 60}, 10569 (1999).

\bibitem{dwell}
R. Landauer and T. Martin, Rev. Mod. Phys. {\bf 66}, 217 (1994).

\bibitem{mel}
See e.g. V.I. Mel'nikov, Phys. Rep. {\bf 209}, 1 (1991).

\bibitem{vank}
N.G. van Kampen, {\it Stochastic Processes in Physics and Chemistry}
(North-Holland, Amsterdam, 1992).

\end{thebibliography}
\end{document}